\documentclass[twocolumn]{aastex631}

\usepackage{multirow}
\usepackage{float}
\usepackage{wrapfig}
\usepackage{amssymb}
\usepackage{amsmath, bm}
\usepackage{tabularx} 
\usepackage{array} 

\begin{document}

\title{Enabling Robust Exoplanet Atmospheric Retrievals with Gaussian Processes}
\shorttitle{Enabling Robust Exoplanet Atmospheric Retrievals with Gaussian Processes}

\author[0000-0003-4459-9054]{Yoav Rotman}
\email{Corresponding authors:\\ yrotman@asu.edu \& luis.welbanks@asu.edu}
\affiliation{School of Earth and Space Exploration, Arizona State University, Tempe, AZ, USA}

\author[0000-0003-0156-4564]{Luis Welbanks}
\email{}
\altaffiliation{51 Pegasi b Fellow}
\affiliation{School of Earth and Space Exploration, Arizona State University, Tempe, AZ, USA}

\author[0000-0001-6247-8323]{Michael R. Line}
\affiliation{School of Earth and Space Exploration, Arizona State University, Tempe, AZ, USA}

\author[0000-0002-1052-6749]{Peter McGill}
\affiliation{Space Science Institute, Lawrence Livermore National Laboratory, 7000 East Ave., Livermore, CA 94550, USA}

\author[0000-0002-3328-1203]{Michael Radica} 
\altaffiliation{NSERC Postdoctoral Fellow}
\affiliation{Department of Astronomy \& Astrophysics, University of Chicago, 5640 South Ellis Avenue, Chicago, IL 60637, USA}

\author[0000-0001-8236-5553]{Matthew C.\ Nixon}
\affiliation{Department of Astronomy, University of Maryland, College Park, MD, USA}

\begin{abstract}

Atmospheric retrievals are essential tools for interpreting exoplanet transmission and eclipse spectra, enabling quantitative constraints on the chemical composition, aerosol properties, and thermal structure of planetary atmospheres. The \textit{James Webb Space Telescope} (JWST) offers unprecedented spectral precision, resolution, and wavelength coverage, unlocking transformative insights into the formation, evolution, climate, and potential habitability of planetary systems. However, this opportunity is accompanied by challenges: modeling assumptions and unaccounted-for noise or signal sources can bias retrieval outcomes and their interpretation. To address these limitations, we introduce a Gaussian Process (GP)-aided atmospheric retrieval framework that flexibly accounts for unmodeled features and correlated noise in exoplanet spectra. We validate this method on synthetic JWST observations and show that GP-aided retrievals reduce bias in inferred abundances and better capture model-data mismatches than traditional approaches. We also introduce the concept of mean squared error to quantify the trade-off between bias and variance, arguing that this metric more accurately reflects retrieval performance than bias alone. We then reanalyze the NIRISS/SOSS JWST transmission spectrum of WASP-96~b, finding that GP-aided retrievals yield broader constraints on CO$_2$ and H$_2$O, possibly alleviating tension between previous retrieval results and equilibrium predictions. Our GP framework provides precise and accurate constraints while highlighting regions where models fail to explain the data. As JWST matures and future facilities come online, a deeper understanding of the limitations of both data and models will be essential, and GP-enabled retrievals like the one presented here offer a principled path forward.

\end{abstract}

\section{Introduction} \label{sec:intro}

Over the past three decades, the Hubble Space Telescope (HST), Spitzer Space Telescope, and now the James Webb Space Telescope (JWST) have produced transmission spectra for over two hundred exoplanets \citep[][]{nikolov_trexolists_2022}. These observations offer a powerful probe to study the chemical composition, temperature structure, and properties of aerosols in exoplanet atmospheres \citep{seager_theoretical_2000}. Statistical constraints on these various properties enable insights into the physicochemical processes in exoplanets \citep[e.g.,][]{showman_atmospheric_2020}, planet formation \citep[e.g.,][]{oberg_effects_2011, madhusudhan_co_2012}, and their prospects for habitability \citep[e.g.,][]{catling_exoplanet_2018, seager_search_2018}. 

Estimates of atmospheric properties are typically derived from observed spectra by interpreting them with models of planetary atmospheres. Obtaining statistical constraints on these properties, however, requires Bayesian inference frameworks that couple a parametric model with a sampling algorithm to derive the posterior probability distribution of model parameters given the data; this process is commonly known as an atmospheric retrieval \citep[see e.g.,][for a review]{madhusudhan_atmospheric_2018}. While this method has become ubiquitous in atmospheric characterization studies \citep[e.g.,][]{madhusudhan_temperature_2009, kreidberg_precise_2014, welbanks_mass-metallicity_2019}, the reliability of its estimates depends critically on model assumptions \citep[e.g.,][]{line_influence_2016}, a thorough understanding of degeneracies \citep[e.g.,][]{welbanks_degeneracies_2019}, and the fidelity of the observations \citep[e.g.,][]{benneke_how_2013}.

The start of JWST science operations in 2022 launched exoplanet atmospheric studies into a new era. Its extended wavelength coverage and increased precision have enabled the detection of molecular species previously inaccessible with HST \citep[e.g.,][]{jwst_transiting_exoplanet_community_early_release_science_team_identification_2022, rustamkulov_early_2023, bell_methane_2023, welbanks_high_2024}, as well as unprecedented constraints on the atmospheric composition of smaller and colder planets \citep[e.g.,][]{madhusudhan_carbon-bearing_2023, schlawin_possible_2024}, surpassing previous records set by HST \citep[e.g.,][]{kreidberg_clouds_2014, benneke_water_2019}. However, JWST has also exposed a number of factors that challenge the robustness of spectroscopic interpretations, including instrumental systematics \citep[e.g.,][]{sarkar_exoplanet_2024} and 
spectroscopic uncertainties that are likely underestimated \citep[e.g.,][]{carter_benchmark_2024}, issues that were either largely under control or often overlooked with previous facilities.

The JWST revolution has also reaffirmed earlier lessons about the importance of modeling assumptions in obtaining reliable atmospheric constraints. Previous efforts to move beyond one-dimensional models in the interpretation of spectroscopic data \citep[e.g.,][]{welbanks_atmospheric_2022, espinoza_constraining_2021, nixon_aura-3d_2022} have become increasingly timely, with JWST offering evidence for inhomogeneous terminators \citep[e.g.,][]{espinoza_inhomogeneous_2024, murphy_evidence_2024}. Likewise, studies that considered the impact of heterogeneous stellar photospheres on transmission spectra retrievals \citep[e.g.,][]{pinhas_retrieval_2018, iyer_influence_2020} have proven essential in light of JWST’s unprecedented precision, which has revealed clear signatures of stellar contamination \citep[e.g.,][]{fournier-tondreau_near-infrared_2023, fournier-tondreau_transmission_2024}. Finally, JWST observations have exposed previously unrecognized sources of uncertainty -- phenomena not anticipated by pre-JWST models but now evident through access to new wavelength regimes. These include unidentified molecular absorbers, unmodeled cloud opacities, and other unexpected spectral features \citep[e.g.,][]{tsai_photochemically_2023, powell_sulfur_2024, grant_jwst_2023, dyrek_so2_2024, welbanks_high_2024}.

With the rapidly growing number of exoplanet spectra now being acquired, the need for reliable atmospheric constraints, despite imperfect data and incomplete models, has become more pressing than ever. The field requires a framework capable of delivering constraints that are both precise and accurate, while robustly marginalizing over uncertainties arising from model assumptions and observational limitations. These inferences must reflect not only what we know, but also what we do not. In this work, we set out to develop and validate such a framework, drawing on advanced statistical methodologies and lessons from other areas of astrophysics where model–data discrepancies have long been recognized and addressed. Non-parametric approaches, in particular, have been used in other areas of astrophysics to mitigate model incompleteness, such as in the interpretation of stellar and brown dwarf spectra, where interpolated forward models often fail to capture observed features \citep[e.g.,][]{czekala_constructing_2015, zhang_uniform_2021, iyer_sphinx_2023}.

Motivated by the need to address model–data discrepancies in retrievals, we turn to Gaussian processes \citep[GPs; e.g.,][]{rasmussen_gaussian_2005}, a non-parametric modeling technique widely used in data-driven analyses for their flexibility and capacity to model complex structure without relying on explicit functional forms. Their application in exoplanetary science has improved light-curve fitting \citep[e.g.,][]{gibson_gaussian_2012, barros_improving_2020, radica_2024_muted}, mitigation of stellar activity in radial velocity searches \citep[e.g.,][]{rajpaul_gaussian_2015, cloutier_2019_characterization}, and telluric signal removal in high-resolution ground-based spectroscopy \citep{meech_applications_2022}.

In this work, we set out to advance the atmospheric retrieval paradigm by revisiting the assumptions embedded in the likelihood function and introducing a more flexible noise model based on Gaussian processes. In what follows, we describe our methodology in Section~\ref{sec:methods}. We benchmark our framework in Section~\ref{sec:tests} using synthetic observations, comparing its performance against traditional retrieval approaches. In the same section, we introduce the concept of the bias–variance tradeoff and demonstrate its usefulness in contextualizing retrieval performance, beyond the more commonly used notion of bias alone. In Section~\ref{sec:W96}, we apply our GP-based retrieval framework to JWST Early Release Observation (ERO) data of the hot Saturn WASP-96~b. We conclude with a summary of our findings and a discussion of their implications in Section~\ref{sec:discussion}.

\section{Methodology} \label{sec:methods}

Our retrieval framework combines a forward model of the planetary atmosphere with a Bayesian inference scheme to derive statistical constraints on atmospheric parameters from transmission spectra. In this work, we build upon the CHIMERA retrieval framework \citep[e.g.,][]{line_systematic_2013, mai_exploring_2019}, extending it with a non-parametric noise model based on Gaussian Processes to better account for model–data mismatches. We draw on previous implementations of \texttt{Celerite} \citep{celerite} and \texttt{George} \citep{george} for modeling covariance functions and evaluating GP likelihoods in \textsl{Aurora} \citep{welbanks_aurora_2021}, but describe our approach in a general, framework-independent manner to promote broad applicability to other retrieval tools. In the following sections, we briefly describe the components of the atmospheric model and the considerations involved in implementing GPs into the retrieval framework.

\subsection{Forward Model}\label{sec:retrieval}

We compute the transmission spectrum of a planet in transit using the atmospheric model from CHIMERA \citep[e.g.,][]{line_systematic_2013, mai_exploring_2019}. Broadly, our atmospheric model solves radiative transfer for a one-dimensional plane-parallel atmosphere in hydrostatic equilibrium. It calculates the transmittance of the atmosphere layer-by-layer, at each wavelength, and integrates along the entire annulus of the planet. Here, we adopt a pressure grid of 100 layers in log-uniform spacing from $10^{-8.7}$ to $10^{1.2}$ bar.

The chemical composition of the atmosphere is parameterized using independent free parameters for the volume mixing ratios of each included molecule or atom, assumed to be constant with height. The model considers opacity from H$_2$O \citep{polyansky_exomol_2018}, CO \citep{li_rovibrational_2015}, CO$_2$ \citep{huang_isotopic-independent_2012}, CH$_4$ \citep{hargreaves_accurate_2020}, NH$_3$ \citep{coles_exomol_2019}, H$_2$S \citep{azzam_exomol_2016}, SO$_2$ \citep{underwood_exomol_2016}, as well as Na and K \citep{allard_k-h2_2016,allard_temperature_2023}, and H$_2$-H$_2$ and H$_2$-He collision induced absorption \citep{karman_update_2019}. 

The vertical temperature structure is parameterized following the prescription of \citet{madhusudhan_temperature_2009}. Inhomogeneous clouds and hazes (aerosols) are modeled using an adaptation of the single-sector model from \citet{welbanks_aurora_2021} where instead of a cloud-top pressure, we consider a grey cloud with an opacity of $\kappa_{\rm cloud}$. The contribution from hazes is modeled as a modification to Rayleigh-scattering with a slope of $\gamma$ and enhancement $a$. The inhomogeneous cloud/haze cover is modeled as a linear combination of a clear and cloudy/hazy atmosphere following \citet{line_influence_2016} where $\phi_{\rm cloud}$ is a free parameter describing the cloud/haze fraction. Finally, we include a scaling parameter, $\times$R$_{\text{pl}}$, that accounts for uncertainty in the reported radius of the planet and its degeneracy with the reference pressure \citep{welbanks_degeneracies_2019}. The spectral model is evaluated at a resolution of $R=100,000$ before being binned to the resolution of the observations. The Bayesian inference is performed using the \texttt{MultiNest} nested sampling algorithm \citep{feroz_multinest_2009} via \texttt{PyMultiNest} \citep{buchner_x-ray_2014}.

\subsection{Retrieval Framework} \label{subsec:covariance}

Although atmospheric retrievals in exoplanetary science originated from grid-based searches \citep[e.g.,][]{madhusudhan_temperature_2009-1}, they are now largely synonymous with Bayesian inference. The widespread availability and adoption of Bayesian sampling methods (e.g., MCMC, nested sampling) has been instrumental in the growth of retrieval frameworks over the past two decades. At the heart of this approach lies Bayes' theorem,

\begin{equation}
    {P(\theta_\mathcal{M}|\mathcal{D}, \mathcal{M}) = \frac{P(\theta_\mathcal{M}|\mathcal{M})P(\mathcal{D}|\theta_\mathcal{M}, \mathcal{M})}{P(\mathcal{D}| \mathcal{M})}}  
\end{equation}

\noindent which expresses the posterior probability distribution $P(\theta_\mathcal{M}|\mathcal{D}, \mathcal{M})$ of the model parameters given the data and modeling assumptions. The marginalized likelihood, or evidence, $\mathcal{Z} = P(\mathcal{D}| \mathcal{M})$, is the likelihood integrated over the prior volume; it serves as a normalization constant and is commonly used for model comparison \citep[see e.g.,][]{welbanks_application_2023}. The prior $\Pi = P(\theta_\mathcal{M}|\mathcal{M})$ represents our initial degree of belief in the model parameters before incorporating the data. The likelihood $\mathcal{L} = P(\mathcal{D}|\theta_\mathcal{M}, \mathcal{M})$ describes the probability of obtaining the observed data given a particular model and parameter set. This term is especially critical, as it encodes the assumed data-generating process and noise properties, and thus defines how differences between the model and data are interpreted, substantially influencing the resulting parameter estimates.

\begin{figure}
    \centering
    \includegraphics[width=0.5\textwidth]{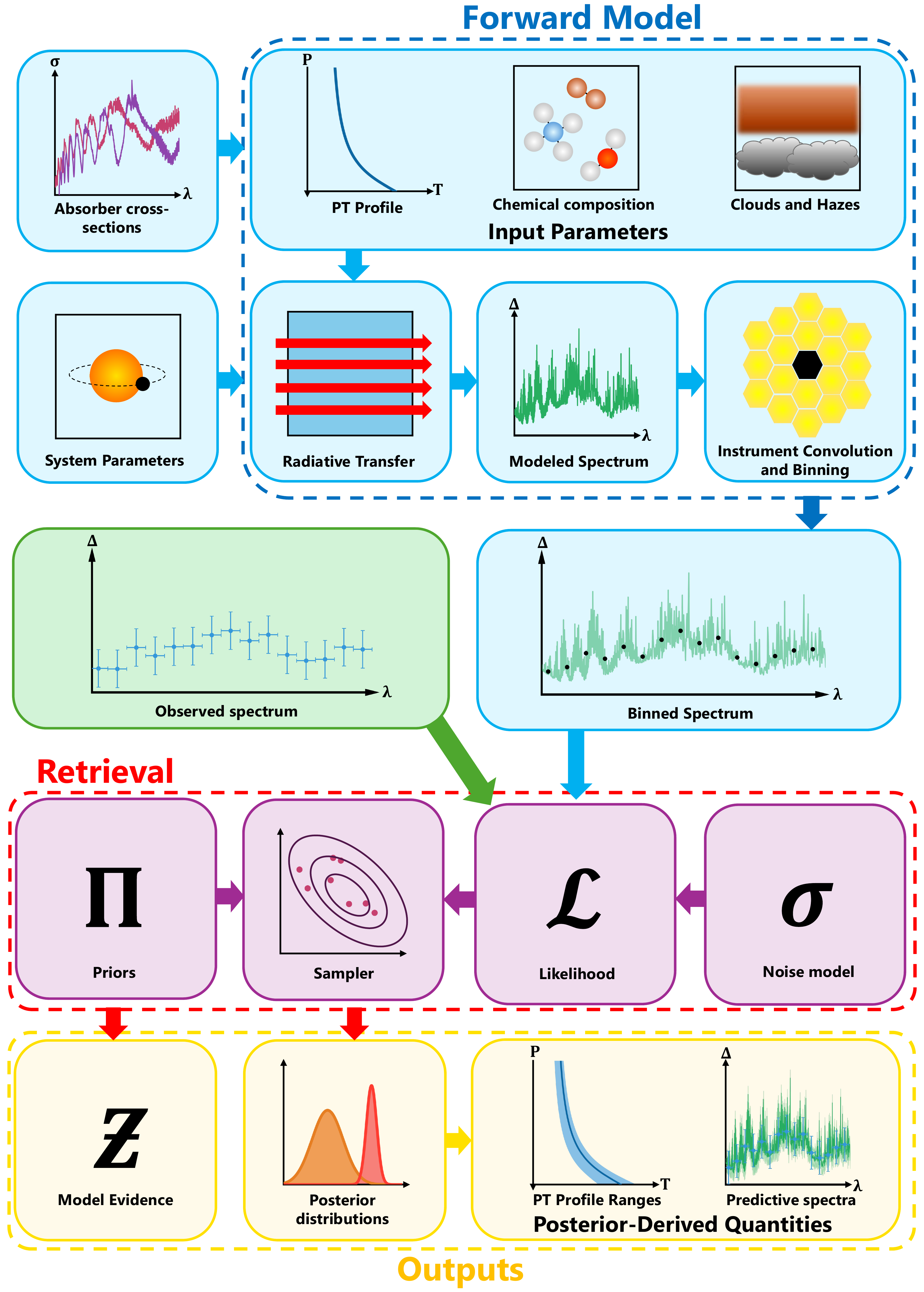}
    \caption{A schematic of our retrieval framework. The forward model generates model evaluations, which are compared to the data through the likelihood. The retrieval component of the framework refers to the Bayesian tools which enable the parameter estimation: a sampling algorithm, parameter priors, and a likelihood function. Our method advances traditional retrieval frameworks by including a Gaussian process noise model.} \label{fig:retrievals_how_do_they_work}
\end{figure}

Traditionally, retrieval frameworks assume that the observational uncertainties are independently and normally distributed, leading to the use of a one-dimensional Gaussian likelihood function, often expressed in log form, to quantify the goodness-of-fit to exoplanet spectra

\begin{equation} \label{eq:chisq}
\ln(\mathcal{L}) = -\frac{1}{2}\sum_{i=1}^n \left[\left(\frac{\bm{{d}}_i - \bm{{m}}_i(\lambda)}{\sigma_i}\right)^2 + \ln(2 \pi \sigma_i^2)\right]
\end{equation}

\noindent where \textit{n} is the number of wavelength bins, ${\bm{d}}_i$ is the observed flux (or transit depth) in the $i^\text{th}$ bin with associated uncertainty $\sigma_i$, and ${\bm{m}}_i(\lambda)$ is the model prediction at that wavelength. In this form, the second term of the equation is a constant penalty term representing the white noise of the data.

While the assumption of independent, normally distributed uncertainties often holds reasonably well in low-resolution or low-precision datasets where noise is typically dominated by photon noise and correlations are less apparent, it intrinsically assumes that all data points and their uncertainties are independent, and that the noise is Gaussian \citep{andrae_dos_2010}. This assumption breaks down in the presence of underlying correlations, which can arise not only from instrumental systematics and data reduction choices \citep[e.g.,][]{schlawin_jwst_2020, ih_understanding_2021, holmberg_exoplanet_2023}, but also from missing or incomplete physics in the forward model. In the latter case, residuals may exhibit structured deviations where the model fails to fully capture the observed spectrum, resulting in correlated noise that violates the standard likelihood assumptions.

To account for these correlations, we include a Gaussian process (GP) in our retrieval framework (Figure~\ref{fig:retrievals_how_do_they_work}). A GP is defined by two components: a mean function, $\mathbf{m}{(\lambda)}$, which in our case corresponds to the modeled atmospheric spectrum, and a covariance matrix, $\mathbf{C}$, which captures correlations between any two input points and is wholly defined by one or more parametric kernel functions $K$ that are defined over an infinite domain. These kernels are each defined by their associated hyperparameters \citep{rasmussen_gaussian_2005}. 

\begin{figure*}
    \centering
    \includegraphics[width=\textwidth]{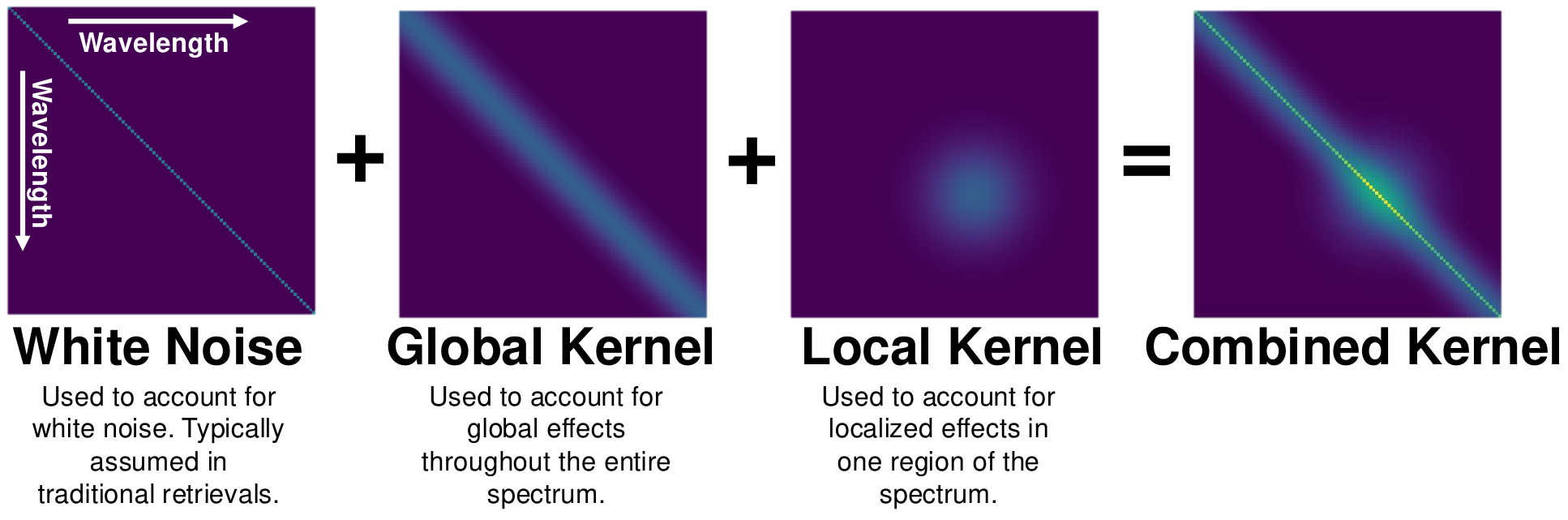}
    \caption{Examples of Gaussian Process (GP) kernels used in our retrieval framework. Traditional retrievals assume uncorrelated white noise, resulting in a diagonal covariance matrix (left). Our approach includes a global kernel that captures long-range correlations across the spectrum (middle), and local kernels that identify localized regions of high correlation (right).} \label{fig:covariance}
\end{figure*}

Each element in the matrix $\mathbf{C}_{ij}$ quantifies the covariance between data points $i$ and $j$, while the diagonal elements $\mathbf{C}_{ii}$ additionally represent the white noise variance at each wavelength bin. When noise is uncorrelated and well-characterized, the off-diagonal terms vanish, and the standard Gaussian likelihood is recovered. A global kernel captures long-range correlations across the entire spectrum, whereas local kernels can be used to model confined regions exhibiting strongly correlated residuals. Figure~\ref{fig:covariance} shows a visual example of a covariance matrix that includes both global and local kernels.

We incorporate this covariance matrix directly into the likelihood function \citep[e.g.,][]{gibson_gaussian_2012, czekala_constructing_2015}. In this case, Equation~\ref{eq:chisq} becomes

\begin{equation} \label{eq:loglike}
    \ln(\mathcal{L}) = -\frac{1}{2} \left[ \bm{r}^\text{T} \mathbf{C}^{-1} \bm{r} + \ln(\det \mathbf{C}) + n \ln 2\pi \right]
\end{equation}

\noindent where $\bm{r} = \bm{d} - \bm{m}(\lambda)$ is the residual between the observed spectrum and the binned model evaluation. $\mathbf{C}$ is the GP covariance matrix, which consists of the covariance kernel $K(\lambda_i,\lambda_j)$ evaluation at each pair of observed wavelengths $\lambda_i$ and $\lambda_j$. In the traditional retrieval case, where data points are assumed to be uncorrelated, $\mathbf{C}$ reduces to a diagonal matrix with elements $\sigma_i^2$ (Figure~\ref{fig:covariance}). Under this assumption, Equation~\ref{eq:loglike} simplifies to the standard one-dimensional form shown in Equation~\ref{eq:chisq}. Because the GP acts directly on the residuals, it is agnostic to whether discrepancies arise from imperfections in the data or from inadequacies in the model. This allows the retrieval to account for correlated structure in the residuals without requiring explicit identification of its source.

Here, we adopt a covariance matrix composed of both global and local kernels. For the global kernel, we assume a single amplitude and correlation length scale applied uniformly across the spectrum to model correlated noise in the data. Local kernels, in contrast, are fitted to capture correlated structure confined to specific regions. These localized deviations can arise from unmodeled spectral features such as missing absorbers or contamination in the data \citep[e.g.,][]{czekala_constructing_2015}, as well as from localized artifacts introduced during the data reduction process \citep[e.g.,][]{radica_awesome_2023, holmberg_exoplanet_2023}. Accounting for these features with local kernels allows the retrieval to marginalize over regional model–data mismatches that might otherwise bias inferred atmospheric properties. Notably, such biases may not be evident in traditional retrieval frameworks, as residuals in one part of the spectrum can still bias the retrieval outcome for absorbers with features elsewhere in the spectrum \citep[see e.g.,][]{welbanks_application_2023}. In addition, the retrieved hyperparameters of the local kernels can help identify which regions of the spectrum contribute most to poor fits and quantify the extent of the correlated deviations.

The covariance kernels are combined additively:

\begin{equation} \label{eq:covariance}
    {K} = k_{\rm trad} + k_{\rm glob} + \sum_{i=1}^n k_{{\rm \ell}, i}
\end{equation}

\noindent where $k_{\rm trad} = \sigma_i^2$ represents the traditional diagonal covariance matrix, $k_{\rm glob}$ is the global kernel, and $k_{{\rm \ell}, i}$ is the $i^{\text{th}}$ local kernel. The combined kernel, $K(\lambda_i, \lambda_j)$, is evaluated at each pair of observed wavelengths to form the covariance matrix $\mathbf{C}$ of the data.

\subsection{Kernel Selection} \label{sec:global}

The global kernel is defined by a characteristic amplitude $a_G$ and length scale $L_G$, where the length scale represents the distance over which wavelength bins exhibit non-negligible correlation \citep{rasmussen_gaussian_2005}. Typically, a positive semi-definite kernel, i.e., one for which $K_{ij} \geq 0$ for all $i, j$, is applied across the entire mean function. In this work, we model the global covariance using a squared exponential (SE) kernel,

\begin{equation} \label{eq:squexp}
    k_{\rm glob} = a_G^2 \text{ exp}\left( \frac{-(\Delta\lambda)^2}{2L_G^2} \right)
\end{equation}

\noindent where $\Delta\lambda_{ij}$ is the distance between any two wavelength bins $i$ and $j$. This kernel has been used across astrophysical literature, including in \citet{gibson_gaussian_2012} and \citet{rajpaul_gaussian_2015}. Later studies, including \citet{czekala_constructing_2015} and \citet{meech_applications_2022}, have used a Matérn-3/2 kernel, which can be considered a rougher variant of the SE kernel. We compare the two on the synthetic spectrum in Section~\ref{sec:tests} and find little difference in performance, as both are able to reliably constrain the injected correlated noise and provide accurate inferences on absorber abundances. In this work, we focus specifically on observations obtained with NIRISS/SOSS, which has a point spread function (PSF) oversampled by 2--3 pixels along the dispersion axis \citep{albert_near_2023}. Under these conditions, a Gaussian approximation is well-suited to model the covariance between adjacent pixels. We therefore choose to adopt the SE kernel throughout this work for consistency between kernels.

In addition to a global covariance, localized regions of high residuals can be modeled using local kernels. We assume the $n^{\rm th}$ local kernel to be a Gaussian function (SE kernel) with amplitude $a_n$, length scale $l_n$, and central wavelength $\mu_n$, such that

\begin{equation} \label{eq:squexp_loc}
    k_{\ell, n} = a_n^2 \, \exp\left( \frac{-(\mu_i - \mu_n)^2}{2L_n^2} \right).
\end{equation}

\noindent The SE kernel acts as a smooth, low-resolution approximation to unresolved or unmodeled spectral features. While it can effectively account for unknown absorbers or contamination, its functional form is not flexible enough to mimic high-resolution line shapes. This makes it unlikely to compete with physically motivated opacity sources when included in the model, ensuring that the GP captures residual structure without affecting the detection or characterization of real atmospheric features.

\begin{figure*}
    \centering
    \includegraphics[width=\textwidth]{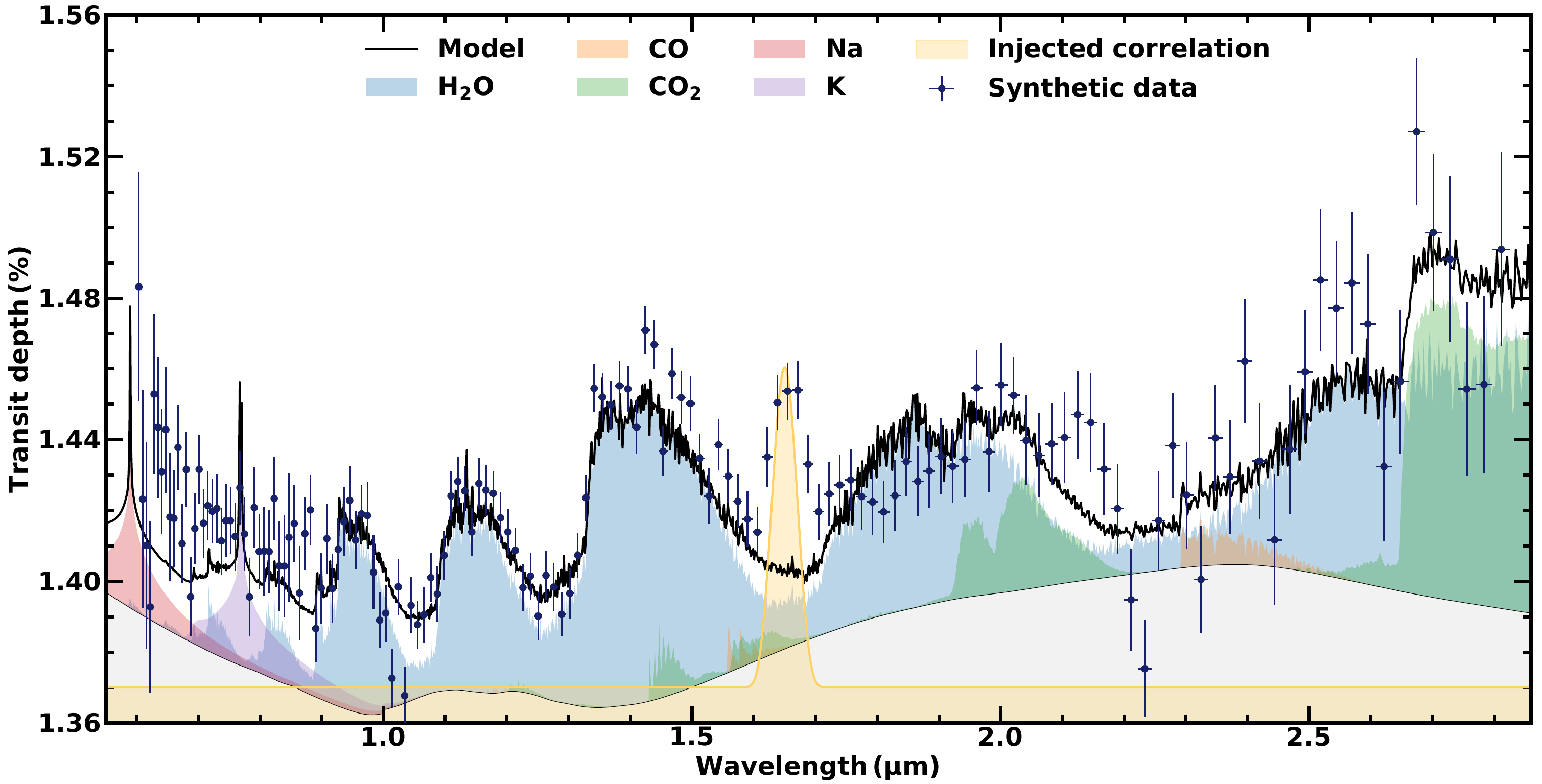}
    \caption{Synthetic transmission spectrum of WASP-96~b used to validate the GP-enabled retrieval framework. The forward model (solid line) is generated using the median atmospheric parameters reported by \citet{taylor_awesome_2023}, and the synthetic data adopt the wavelength bin widths and precision from \citet{radica_awesome_2023}. Correlated noise is injected across the spectrum, including a localized feature centered at 1.65~$\mu$m (yellow shaded region), to mimic unmodeled structure (e.g., unidentified absorber). The resulting synthetic data points (blue markers) exhibit realistic observational scatter while preserving the imposed correlation structure. Colored regions indicate the contributions of key opacity sources to the model spectrum.}
    \label{fig:toy_model}
\end{figure*}

\section{Validation on Synthetic NIRISS/SOSS Spectra}\label{sec:tests}

We validate our GP-enabled retrieval framework on synthetic JWST/NIRISS SOSS observations of the hot Saturn WASP-96~b \citep{hellier_transiting_2014}. The model used to generate the simulated spectrum, spanning 0.5–3.4~$\mu$m, was computed at a resolution of $R = 100{,}000$, with input parameters for the chemical composition and temperature profile corresponding to the median values reported by \citet{taylor_awesome_2023}. We include a half-cloudy terminator ($\phi_{\rm cloud} = 0.5$), with a grey cloud opacity of $\kappa_{\rm cloud} = 10^{-31}$~m$^2$ and an enhanced scattering slope.

The resulting atmospheric model was then binned to the wavelengths and widths from \citet{radica_awesome_2023}. In addition to white noise drawn from the reported precision of the JWST/NIRISS SOSS spectrum in \citet{radica_awesome_2023}, we inject global correlated noise across the entire wavelength range, with an amplitude of $a_G = 100$~ppm and a correlation length scale of $L_G = 0.05$~$\mu$m\. We also inject a localized high-correlation feature centered at 1.65~$\mu$m. The final synthetic spectrum is generated as a single random draw from a multivariate normal distribution, incorporating both white and correlated noise. This approach simulates realistic observational scatter while preserving the imposed correlation structure. In doing so, we test whether the retrieved hyperparameters can inform inferences about the underlying correlation structure and the specific wavelengths most affected. The model, injected noise, and resulting synthetic data are shown in Figure~\ref{fig:toy_model}. In addition to the noticeable feature at $\sim$1.65~$\mu$m, the simulated noise results in a generally higher transit depth relative to the input model at wavelengths $\lambda \lesssim 1$~$\mu$m.

We compare four retrieval configurations to evaluate the impact of different covariance structures. The first is a ``traditional'' retrieval, which assumes uncorrelated Gaussian noise and adopts a diagonal covariance matrix: $\mathbf{C} = \text{diag}(\sigma^2)$. The second is a ``local-only'' retrieval, which adds a single local kernel to account for a region of high data–model mismatch: $\mathbf{C} = \text{diag}(\sigma^2) + k_\ell(a_1, L_1, \mu_1)$. The third is a ``global-only'' retrieval, where a global kernel is used to capture long-range correlations across the spectrum: $\mathbf{C} = \text{diag}(\sigma^2) + k_g(a_G, L_G)$. Finally, we test a ``combined'' retrieval that includes both global and local kernels:

\begin{equation} \label{eq:stack_cov}
    \mathbf{C} = \text{diag}(\sigma^2) + k_G(a_G, L_G) + k_\ell(a_1, L_1, \mu_1),
\end{equation}

\noindent which corresponds to the specific case of Equation~\ref{eq:covariance} with one local kernel.  

Table~\ref{table:W96_prior} provides the priors for the parameters and hyperparameters considered. We elect to sample along uniform priors for each hyperparameter, rather than log-uniform priors, as this provides the most informative hyperparameter posterior distributions. We motivate this choice further in Section \ref{sec:hyperparams_discussion}. However, we caution that prior selection should be done on a case-by-case basis, and that improper choice of prior can lead to biasing effects in the posterior distribution, particularly when the expected values for a parameter cover multiple orders of magnitude.


\begin{table}
\centering
\caption{Priors for retrieval parameters and GP hyperparameters used in synthetic and observed spectra fits for WASP-96~b. $\mathcal{U}$ denotes a uniform prior.}
\label{table:W96_prior}
\renewcommand{\arraystretch}{1.1}
\begin{tabular}{ll}
  \textbf{Parameter} & \textbf{Prior} \\
  \hline
$\log_{10}(X_i)$ & $\mathcal{U}[-12, -0.3]$ \\
xR$_{\rm pl}$ & $\mathcal{U}[0.5, 1.5]$ \\
$\log_{10}(\kappa_{\rm cloud})$ [m$^2$] & $\mathcal{U}[-40, -25]$ \\
$\phi_{\rm cloud}$ & $\mathcal{U}[0, 1]$ \\
$\log_{10}(a)$ & $\mathcal{U}[-4, 10]$ \\
$\gamma$ & $\mathcal{U}[-2, 20]$ \\
$T_0$ [K] & $\mathcal{U}[800, 1400]$ \\
$\log_{10}(P_{1,2})$ [bar] & $\mathcal{U}[-8.7, 1.2]$ \\
$\log_{10}(P_3)$ [bar] & $\mathcal{U}[-2, 1.2]$ \\
$\alpha_{1,2}$ [K$^{-1/2}$] & $\mathcal{U}[0.02, 2]$ \\
\hline
\multicolumn{2}{c}{\textbf{GP Hyperparameters}} \\
\hline
$a_G$ [ppm] & $\mathcal{U}[0, 10^4]$ \\
$L_G$ [$\mu$m] & $\mathcal{U}[0, 1]$ \\
$\mu_n$ [$\mu$m] & $\mathcal{U}[0.05, 3.6]$ \\
$a_n$ [ppm] & $\mathcal{U}[0, 10^4]$ \\
$L_n$ [$\mu$m] & $\mathcal{U}[0, 0.2]$ \\
\end{tabular}
\end{table}

\subsection{Retrieval Performance on Synthetic WASP-96~b Data}\label{sec:toy_results}

\begin{figure*}
    \centering
    \includegraphics[width=\textwidth]{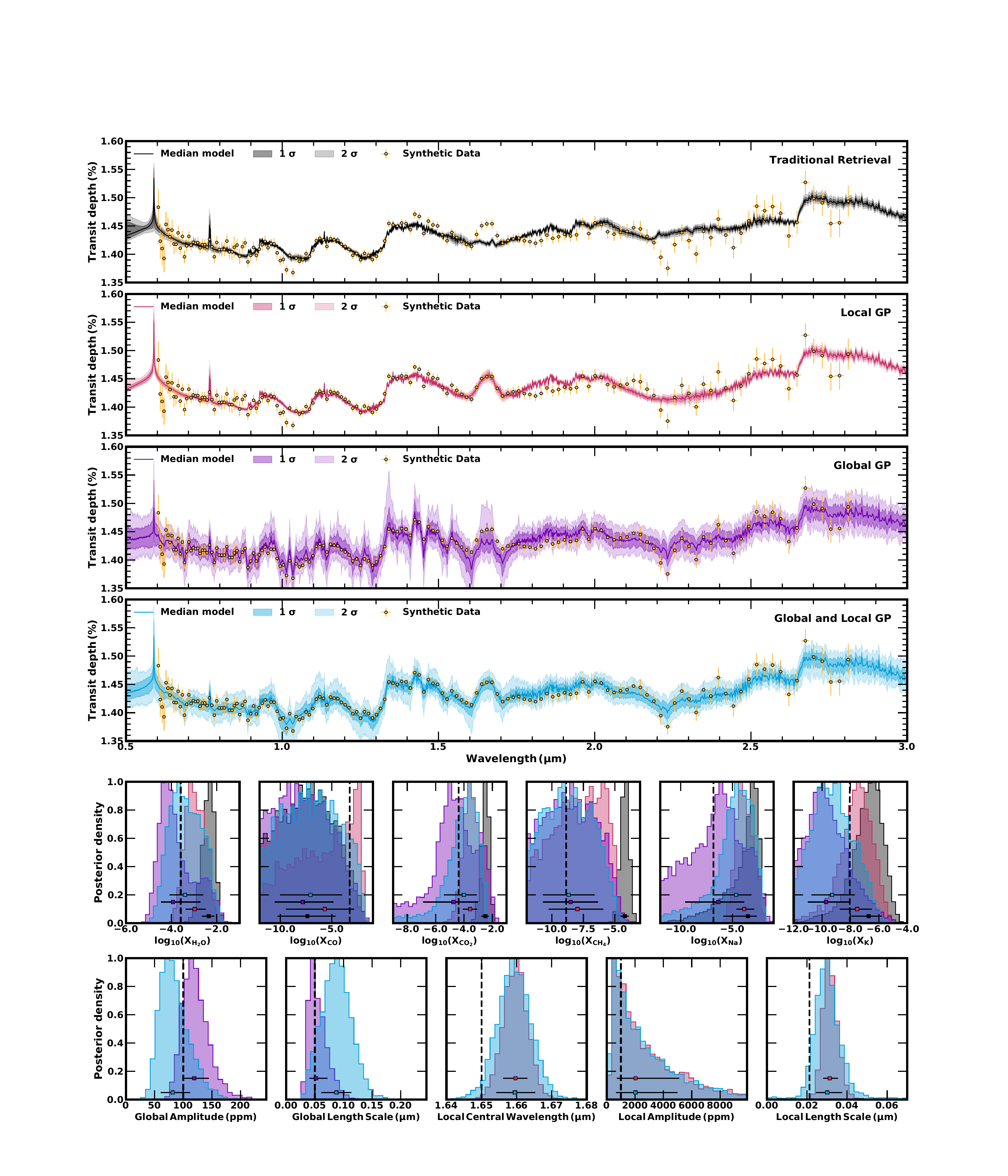}
    \caption{The retrieved median spectra for each of the four retrieval frameworks with the simulated data (top), alongside the posterior distributions for the H$_2$O, CO, CO$_2$, CH$_4$, Na, and K abundances (middle) and the GP hyperparameters (bottom). While the traditional framework provides precise constraints, it overestimates the H$_2$O, CO$_2$, Na, and K abundances by significant margins; this is somewhat, but not entirely, rectified by the inclusion of a local kernel. Conversely, a global kernel provides accurate constraints on all four, but with larger, uninformative posterior distributions. By combining the global and local kernels, the precision of the posterior distributions is increased while maintaining accuracy to the true injected abundances (dashed lines). The medians of the hyperparameters are within 3$\sigma$ of the injected values for all three cases, demonstrating the efficacy of this method in retrieving underlying correlated noise in datasets.} \label{fig:toy_results}
\end{figure*}

Although the traditional retrieval produces the most precise posterior distributions, it yields highly inaccurate results due to the injected correlated noise. The median retrieved abundances of H$_2$O, CO$_2$, and CH$_4$ are offset from the true values by 1.24, 1.89, and 4.64 dex, respectively (Figure~\ref{fig:toy_results}), demonstrating general overestimates due to unmodeled noise effects, particularly for CH$_4$ abundance.

In the ``local-only'' retrieval, a single local kernel is used to capture a high-correlation feature in the spectrum. This configuration enables the retrieval to constrain the injected feature’s amplitude and central wavelength, but it does not account for the global correlations present across the spectrum. As a result, the retrieval remains biased in key atmospheric parameters: the H$_2$O and CO$_2$ abundances are still offset by 0.62 and 0.81 dex, respectively. However, the local kernel successfully mitigates the effect of the injected feature near 1.7~$\mu$m, which overlaps with a CH$_4$ absorption band, resulting in a CH$_4$ abundance much closer to the true value than in the traditional case (0.89 dex offset rather than 4.64 dex). This suggests that while local kernels can improve model–data agreement in targeted regions, they are insufficient for capturing broader correlated structure.

In contrast to the traditional case, the retrieval using a global kernel recovers all gas abundances to within 2$\sigma$ of their true values, but at the expense of precision. For example, the CO$_2$ abundance is now much more accurate (0.36 dex offset compared to 1.89 dex in the traditional case), but the 1$\sigma$ posterior interval spans nearly 4 dex (Figure~\ref{fig:toy_results}). As a result, the retrieval provides only weak constraints that are insufficient to inform parameters such as atmospheric metallicity or the carbon-to-oxygen ratio.

Finally, we consider the retrieval that includes both global and local kernels. This configuration enables the GP to accurately constrain all hyperparameters (Figure~\ref{fig:toy_results}, bottom), including successful identification of the injected feature at 1.65~$\mu$m. It also yields atmospheric parameters that are much closer to the true values, without substantially increasing the posterior variance. The retrieved median abundances of H$_2$O, CO$_2$, and CH$_4$ differ from their true values by only 0.19, 0.37, and 
0.20 dex, respectively (Figure~\ref{fig:toy_results}). The Na abundance is measured with higher precision than in the global-only case and greater accuracy than in the traditional and local-only retrievals, although it is still overestimated relative to the input. This overestimation may be partially attributable to the specific noise realization in our synthetic data, which diverges from the input model at $\lambda \lesssim 1$~$\mu$m (Section~\ref{sec:tests}). Additionally, because the NIRISS bandpass captures only the red wing of the Na~I doublet, the sodium abundance is degenerate with the scattering slope, making it sensitive to both the noise realization and the assumed aerosol properties \citep{taylor_awesome_2023}.

Despite its presence in the synthetic data, CO is not significantly detected (a detection significance, e.g., \citet{trotta_bayes_2008}, of 1.1$\sigma$ in the traditional case) and our retrievals can only infer an upper limit, likely due to the lack of a strong absorption feature in the NIRISS bandpass (Figure \ref{fig:toy_model}). Additionally, the abundance of Na is particularly susceptible to bias due to the lack of a complete absorption feature captured by the NIRISS bandpass; only the red tail of the Na I doublet is captured, which leads to a degeneracy with parameters of the aerosol scattering slope \citep{taylor_awesome_2023}.

We assess the performance of these different models with a model comparison based on the Bayes factor and its conversion to a `detection significance' \citep[e.g., model preference, ][]{trotta_bayes_2008,benneke_how_2013,welbanks_aurora_2021}. We find that the global-only and local-only GP frameworks are preferred over the traditional framework by $8.4\sigma$ (log-Bayes factor $=33.65$) and $7.8\sigma$ ($\ln B= 28.41$) respectively. Likewise, the combined local and global GP is preferred over the traditional framework by $8.2\sigma$ ($\ln B= 31.96$). All cases correspond to a strong indication of the need for a GP \citep{trotta_bayes_2008} but no statistically significant preference regarding the use of a local kernel when a global kernel is also used ($2.4\sigma$, $\ln B = 1.69$). We attribute this to the small amplitude of the injected feature, allowing a global GP to account for it without the need for a local kernel (Figure \ref{fig:toy_results}). We consider a case with an higher-amplitude injected feature (Appendix \ref{app:feature}), and find that in the case of the larger feature, the combined global and local GP is preferred over a global GP by $7.1\sigma$ ($\ln B=23.29$); this implies that for larger features, global kernels will not be enough to marginalize over localized effects, and that the need for a local kernel is partially correlated with the amplitude of the data-model mismatch.

Additionally, we explore whether a GP will artificially inflate noise for data without underlying correlation, providing a false positive. To test this, we generate identical data to that in Section \ref{sec:tests}, but do not add correlated noise, so that the white noise encodes the full underlying uncertainty in the data. We find that the inclusion of any combination of GP kernels does not significantly impact the retrieved parameters in this scenario; additionally, the traditional framework is preferred over all GP frameworks by $>$3$\sigma$, implying that the GP is not necessary for fitting this dataset accurately. We therefore conclude that our use of a GP-aided retrieval is unlikely to artificially degrade the science output from observations in this work, even in the best-case scenario where a GP is not needed to fully explain the data. This is, however, dependent on both the GP kernel parameterization and expected posterior distribution precisions given a white noise model, among other effects. It should therefore be considered for each case independently. A more complex GP model is more likely to degrade the science output for a given dataset, as model flexibility increases at the cost of precision.

The Bayes factor is used here to indicate the strength of the preference for the retrieval that includes a GP noise model rather than the traditional white noise model. However, it is important to note that by changing the noise model to be more flexible, the GP noise model case can actually be considered a less complex noise model, allowing for structured correlations in the data to be considered within the noise model rather than being attributed to a more complex physical model. The Bayes factor is also highly dependent on the selection of parameter and hyperparameter prior ranges, as well as the assumption that the models being compared are both approximately valid interpretations of the data. Here, we select hyperparameter priors that we expect to be broad enough to not bias the hyperparameters towards a given value, while also remaining within physical bounds (i.e., avoiding correlation length scales larger than the wavelength range of the data or unphysical kernel amplitudes for a transmission spectrum). Bayes factors should therefore not be considered definitive indicators of a need for a GP, and are rather a conditional quantification of the comparison between the GP and non-GP noise models combined with the atmospheric models used in this work. We include further discussion of the selection of hyperparameter priors and their effects in Section \ref{sec:hyperparams_discussion}.

\subsection{Recovery of Injected Noise and Feature}

The inferred hyperparameters of the GP from the retrieval can contain information regarding the underlying noise distribution and any features the model is unable to fit. Here, we compare the injected noise and feature of our dataset to the retrieved hyperparameters to identify how informative they are. We use uniform priors for all hyperparameters. The bottom row in Figure \ref{fig:toy_results} shows the retrieved hyperparameters of the local (pink), global (purple), and the combined global and local GP (blue) frameworks.

In all three cases, we find that the retrieval frameworks are able to provide both precise and accurate estimates of the underlying correlation. Both the global and local kernel amplitudes are consistent with the true amplitudes, with the true values falling within the ``single standard deviation" range of the posterior distribution (16$^{\rm th}$ and 84$^{\rm th}$ percentile). The global length scale is similarly constrained in the global-only case, although it is slightly overestimated in the combined case (with the true value falling in the $7^{\rm th}$ percentile of the posterior distribution, consistent within $2\sigma$ for a Gaussian distribution). Although the local kernel does overestimate the wavelength and width of the injected feature in both the local-only and combined case, these do not significantly affect its ability to retrieve accurate posterior distributions for atmospheric parameters, and are still within a reasonable margin of error away from the injected values (for both, the true value falls in the $6^{\rm th}$ percentile of each posterior distribution). Specifically, the retrieved central wavelength is only $\sim$0.01 $\mu$m from the true value, which we attribute to randomness in the noise instance used to simulate the data and the low resolution of the data. The exact central wavelength may be difficult to constrain for a feature characterized by only $\sim$5 data points, but 0.01 $\mu$m is within a single bin width of the true wavelength. As such, we consider this an accurate inference.

Notably, the global-only case retrieves a higher global correlation amplitude than the combined case. This is due to the absence of a local kernel in the global-only case, leading to overcompensation; to account for the 1.65 $\mu$m feature without a local kernel, the global kernel retrieves a higher amplitude that can marginalize over the feature. Thus, in the combined case, the inferred amplitude is lower, with the true injected value being within $1\sigma$ of the median inferred amplitude.

\subsection{Bias-Variance Tradeoff}
\label{sec:BV}
We now introduce the mean squared error (MSE),

\begin{equation} \label{eq:MSE}
    \text{MSE} = b^2 + \sigma^2
\end{equation}

\noindent where the bias $b$ is the difference between the true value of a parameter and the expectation value (median) of its corresponding posterior distribution, and $\sigma^2$ is the variance of the posterior distribution.
The MSE can thus be used to consider the total error budget and compare which framework best describes the observed spectrum when accounting for both precision and accuracy, both of which are crucial to exoplanet atmospheric inferences. 

\begin{figure*}
    \centering
    \includegraphics[width=\textwidth]{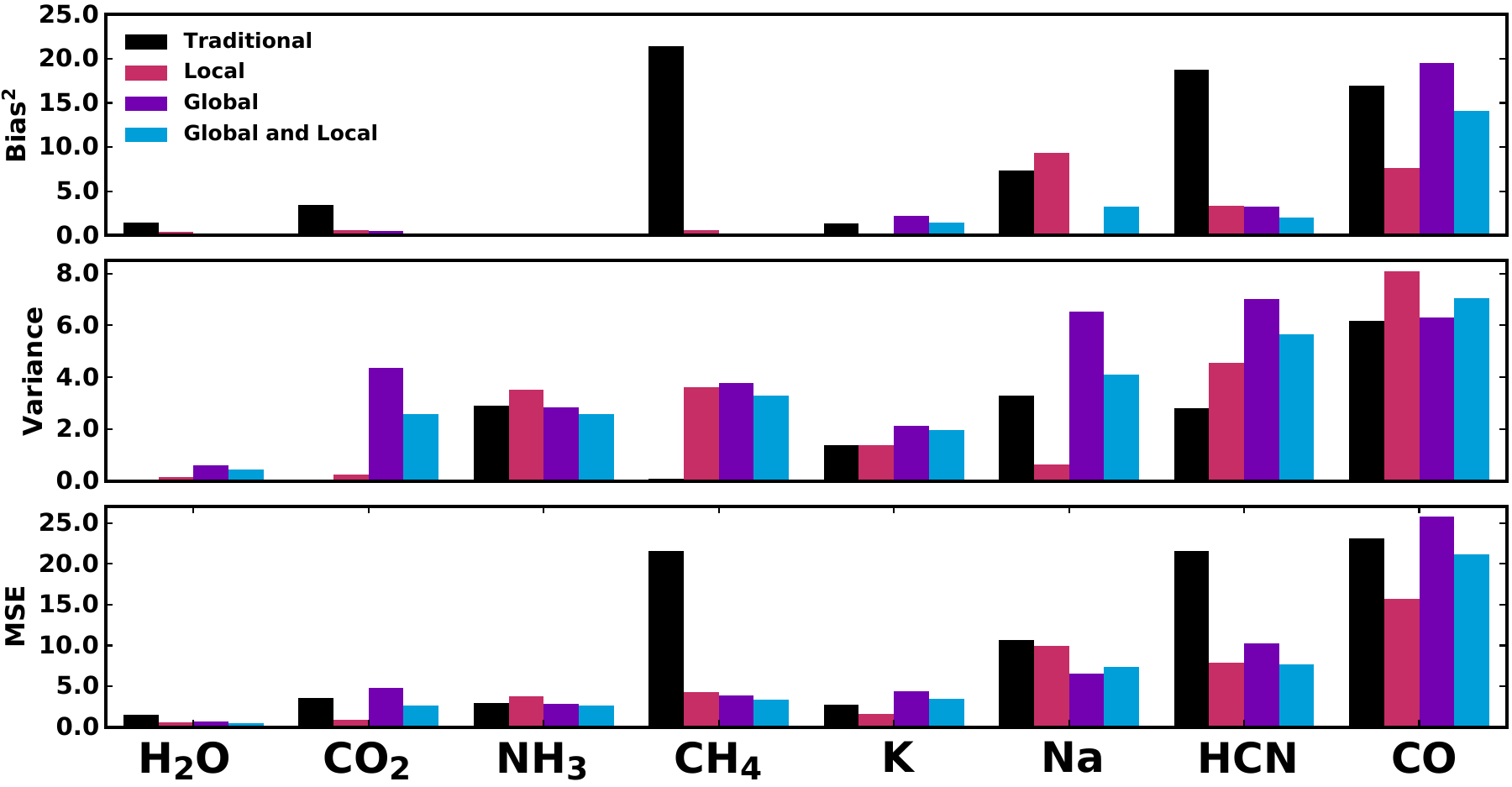}
    \caption{The bias (top), variance (middle), and mean squared error (bottom) of each retrieval framework for the eight abundances used as parameters in our retrieval. The traditional retrieval shows a low variance but a high bias, leading to inaccurate inferences. The global-only and local-only retrieval cases both show low bias but high variance across most parameters to differing extents. A combined global and local retrieval provides the lowest MSE for most of the inferred abundances, and thus the most results with the best combined accuracy and precision.} \label{fig:biasvariance}
\end{figure*}

We compare the MSE of the abundances of the atmospheric absorbers in all three cases (Figure \ref{fig:biasvariance}). The traditional retrieval case shows low variance (high precision), but trades off for a high bias on most of the absorber abundances, resulting in inaccurate retrieved abundances. This is also shown in the local GP case to a lesser extent.  Conversely, the both the local-only and global-only GP cases show a lower bias but a high variance, as the true value is now contained within the posterior distribution for most of the absorbers, but they are wider and less informative. The local-only GP case shows this to a lesser extent, and is able to accurately retrieve abundances for the absorbers with absorption features most impacted by the 1.65 $\mu$m injected feature (e.g., HCN). However, the combined global and local GP scenario shows the lowest MSE for most of the abundances inferred from our dataset, implying a minimized bias-variance trade-off. We thus conclude that a retrieval utilizing combined local and global kernels provides the most informative and accurate posterior distributions for this dataset.   

\section{
Application to the JWST NIRISS/SOSS Spectrum of WASP-96~b}\label{sec:W96}

WASP-96 b is a hot Saturn (M $=0.48$M$_{\text{J}}$, R $=1.2$R$_{\text{J}}$; \citealt{hellier_transiting_2014}) with an equilibrium temperature of $\sim\!1285$ K and an orbital period of 3.4 days. Previous observations of the transmission spectrum of WASP-96 b with the Very Large Telescope (VLT), HST/WFC3, and Spitzer revealed a cloud-free atmosphere with broad Na and K features \citep{nikolov_absolute_2018, yip_compatibility_2020, nikolov_solar--supersolar_2022, mcgruder_access_2022}, making it a prime target for transit observations. However, recent modeling work has called the cloud-free nature of the atmosphere into question \citep{samra_clouds_2023}, and found that a cloudy terminator with efficient vertical mixing may also match the optical/NIR spectra of \citet{nikolov_solar--supersolar_2022}.

\citet{radica_awesome_2023} presented the near-infrared (NIR) transmission spectrum of WASP-96 b observed with JWST NIRISS/SOSS (0.6-2.8 $\mu$m) as part of the JWST Early Release Observations (ERO) program. The spectrum is split into two overlapping SOSS orders (0.6--1.0 $\mu$m and 0.85--2.8 $\mu$m), with an additional third order not considered within the data reduction process. The data was reduced via the \texttt{exoTEDRF} pipeline\footnote{Formerly known as \texttt{supreme-SPOON}} \citep{feinstein_early_2023, radica_awesome_2023, radica_exotedrf_2024}. \citet{radica_awesome_2023} compared the data to a grid of radiative-convective-thermochemical equilibrium models, while \citet{taylor_awesome_2023} performed retrievals with three different frameworks on the spectrum. The results of \citet{radica_awesome_2023}, including a broadened Na feature and a solar to super-solar metallicity, were largely consistent with those of \citet{nikolov_absolute_2018} and \citet{nikolov_solar--supersolar_2022}. However, they found potential for a hazy terminator from an enhanced aerosol scattering slope at wavelengths below $\sim\!1.0$ $\mu$m. This was attributed to a possible degeneracy with the Na abundance, as the NIRISS bandpass contains only a partial Na feature at the blue edge of the spectrum, which could masquerade as a scattering slope \citep{taylor_awesome_2023}. Additionally, \citet{taylor_awesome_2023} inferred a higher CO$_2$ abundance than predicted by the equilibrium models in \citet{radica_awesome_2023}. The latter found best-fit equilibrium models with a 1--5$\times$ solar metallicity and solar C/O ratio, implying a CO$_2$ abundance of $\log_{10}X_{\text{CO}_2} \sim 10^{-7}$. However, the former retrieved a CO$_2$ abundance more broadly consistent with a 10$\times$ solar composition (e.g., $\log_{10}X_{\text{CO}_2} = -4.38^{+0.46}_{-0.57}$ with Aurora). 

The apparent tension between the retrieved chemical composition of WASP-96~b from JWST/NIRISS SOSS and predictions from chemical equilibrium models motivates a re-analysis of these data using our GP-aided retrieval framework. As demonstrated in Section~\ref{sec:toy_results}, this method can provide more realistic expectations for the precision of chemical abundance constraints given a dataset, helping to clarify whether previous CO$_2$ estimates for WASP-96~b reflect overprecision rather than a fundamental disagreement. In addition, this framework can identify localized regions of mismatch between models and data, offering insights that can inform both future observations and modeling efforts.

\begin{figure*}
    \centering
    \includegraphics[width=\textwidth]{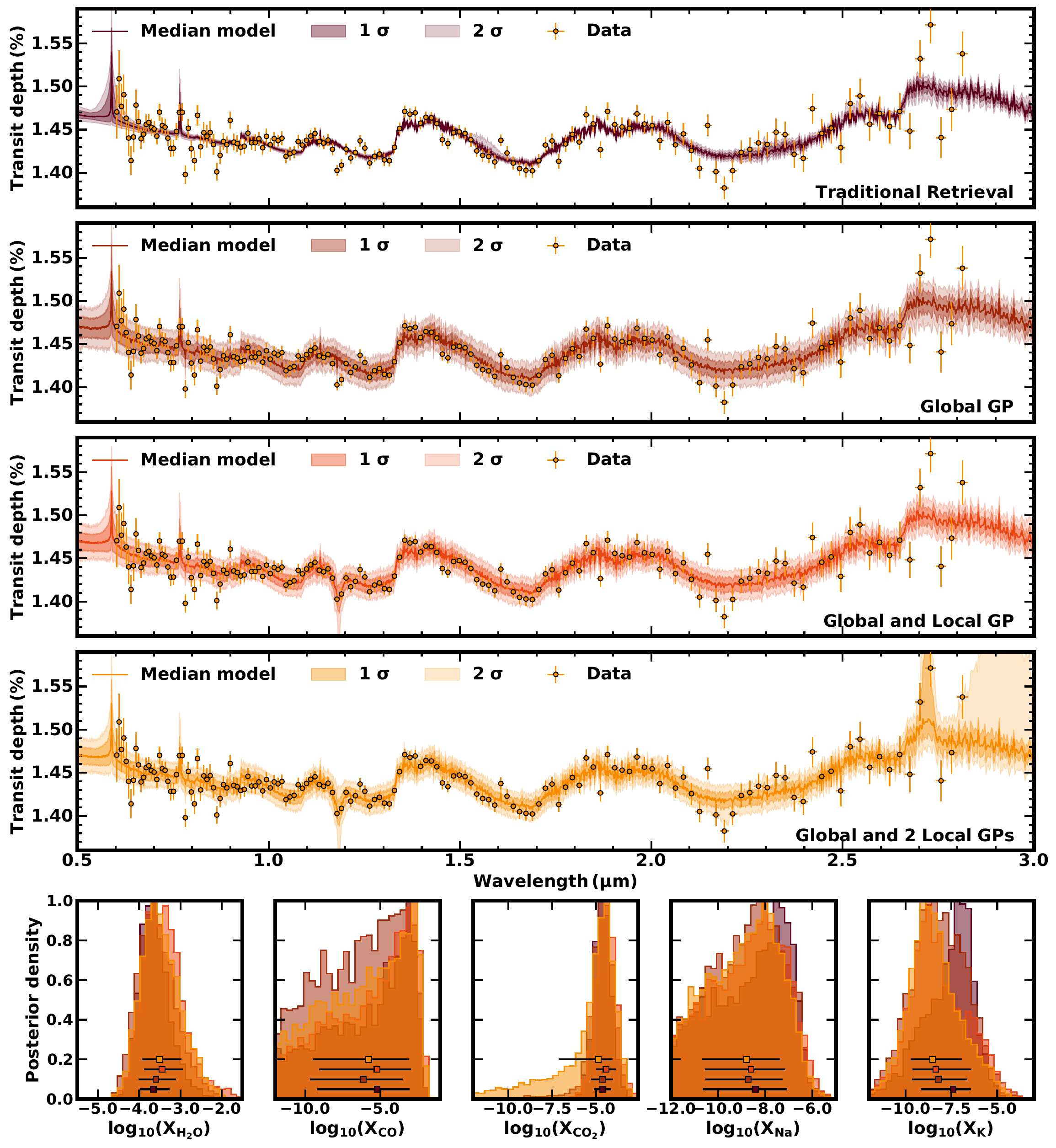}
    \caption{Top: the retrieved spectra, as well as the 1 and 2$\sigma$ confidence intervals, for each retrieval of WASP-96 b (in order; traditional, global-only, global and local, global and two locals). The GP-aided retrievals identify poorly-fit regions at 1.2 and 2.7 $\mu$m, and a small global correlation. Bottom: the retrieved abundances for relevant absorbers in the NIRISS bandpass. The width of the CO$_2$ and H$_2$O abundance posterior distributions increases with the use of a GP, implying that traditional retrievals infer an overly-precise abundance for both species.} \label{fig:spectra1}
\end{figure*}

For this analysis, we consider four retrieval configurations: (1) a traditional retrieval assuming uncorrelated white noise, (2) a GP-aided retrieval with a global kernel, and (3–4) two GP-aided retrievals with one and two local kernels added to the global kernel. When fewer local features are present than kernels allowed, multiple local kernels may overlap in wavelength and stack onto a single region. As the number of local kernels increases, the primary constraint becomes computational rather than statistical, since each kernel introduces additional hyperparameters. In this work, we allow up to two local kernels, each free to explore the full wavelength range. The resulting spectra from each retrieval configuration are shown in Figure~\ref{fig:spectra1}.

\subsection{Revisiting the Atmospheric Composition of WASP-96~b with GP-Aided Retrievals} \label{sec:abundances}

The retrieved parameter values for each retrieval configuration are summarized in Appendix~\ref{app:summary_table}, Table~\ref{table:retrieved}. Key differences in the retrieved abundances are shown in Figure~\ref{fig:spectra1}, while Figure~\ref{fig:hyperparams} shows the posterior distributions for the GP hyperparameters. Overall, the median estimates of the retrieved abundances remain largely consistent across retrieval cases, as expected for well-characterized data. However, for both H$_2$O and CO$_2$ (the two most tightly constrained molecules) the precision decreases in the GP-aided cases. In particular, we find that the true CO$_2$ abundance may be difficult to estimate from the partial coverage of its absorption band in the NIRISS bandpass, resulting in a posterior distribution that is $\sim$1 dex broader than in the traditional retrieval. This is consistent with the usage of a GP noise model, which inherently allows for additional flexibility in the model and thus leads to wider posterior distributions for the model parameters.

The inferred CO$_2$ abundance from the GP noise model retrievals is consistent with both the super-solar metallicity estimate of \citet{taylor_awesome_2023} and the solar metallicity estimate of \citet{radica_awesome_2023}. While the median estimates for H$_2$O abundance remain consistent between retrieval cases, accounting for any underlying correlated noise in the data again leads to a wider posterior distribution (an increase of $\sim\!0.1$ dex). These results imply that the expected precision capability of NIRISS observations on both CO$_2$ and H$_2$O abundance constraints may be less than previously predicted with traditional retrievals. The posterior distributions for CO, Na, and K remain largely consistent across the retrieval cases, implying that the traditional retrieval was able to characterize these accurately and with the true expected precision from NIRISS observations. 

\subsection{Interpreting GP Hyperparameters in WASP-96~b Retrievals} \label{sec:hyperparams}

Across the GP-aided retrievals, there is evidence of a global correlation throughout the entire spectrum with an amplitude of $\sim$60-100 ppm and a length scale of $\sim$0.01 $\mu$m. The retrieved amplitude of the global kernel is higher when a local kernel is not considered (85 ppm) than when a local kernel is considered (70-75 ppm). This is likely due to the global kernel compensating for localized unknown features with a higher amplitude which can be characterized by local kernels in the other cases, similar to the behavior seen in retrievals of the synthetic dataset. Performing a model comparison, we calculate a model preference (e.g., `detection significance') for the inclusion of a GP relative to the traditional retrieval, for all scenarios. Specifically, the Global, Global and 1 Local, and Global and 2 Local cases are preferred over a traditional retrieval by 4.8$\sigma$ ($\ln B = 8.64$), 3.8$\sigma$ (5.78), and 3.8$\sigma$ (5.79), respectively. These results suggest that the inclusion of a GP is strongly preferred over a traditional retrieval \citep{trotta_bayes_2008, benneke_how_2013}.

\begin{figure*}
    \centering
    \includegraphics[width=\textwidth]{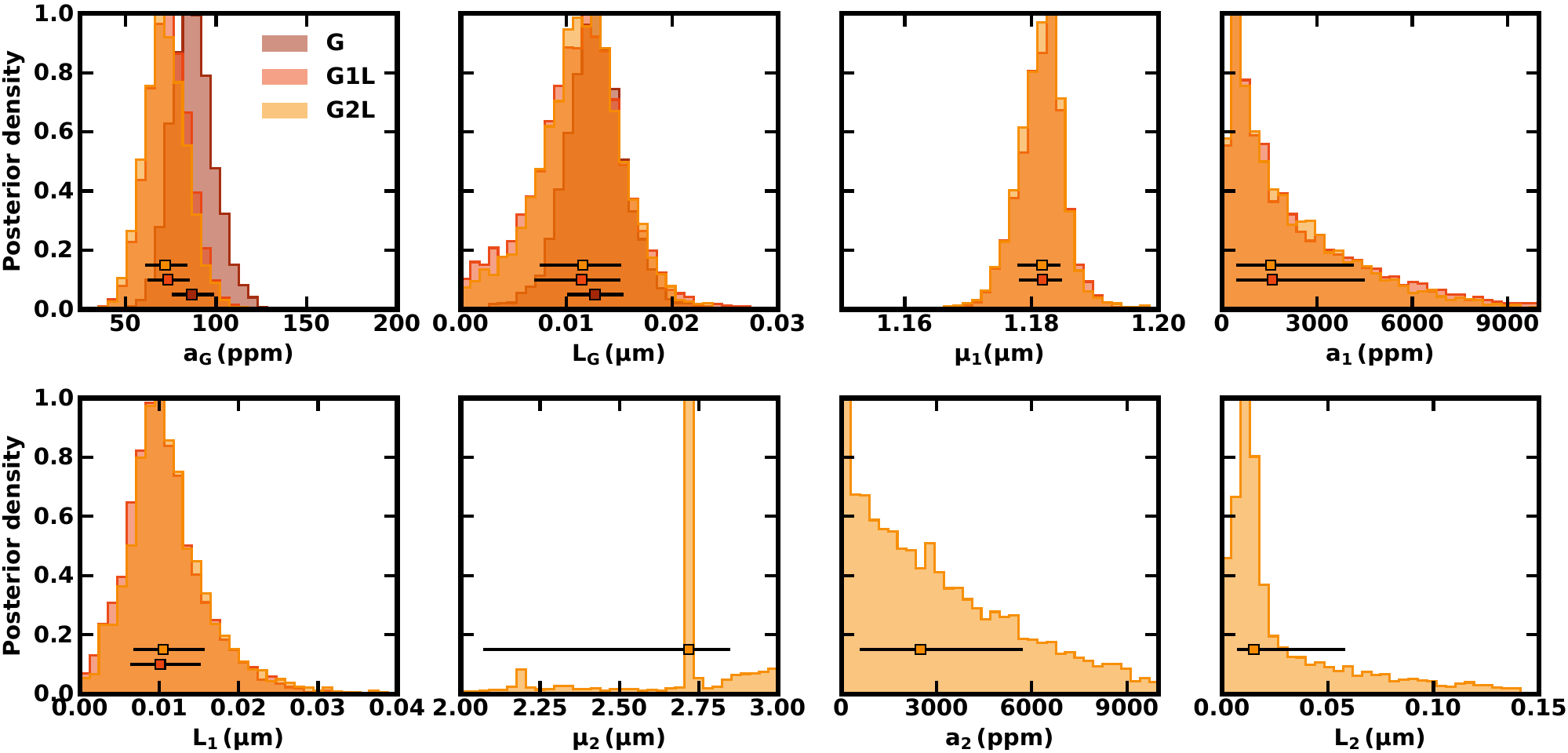}
    \caption{The retrieved hyperparameters for the GP-aided retrievals. Numbered subscripts correspond to hyperparameters of the local kernels, while the "G" subscript corresponds to the global kernel hyperparameters. The hyperparameters of the global kernel imply a correlation of $\sim$60-100 ppm between neighboring wavelength bins, while the local kernels identify regions of data-model mismatch at 1.18 and 2.7 $\mu$m.} \label{fig:hyperparams}
\end{figure*}

However, a global-only kernel is moderately preferred over either combined kernel (i.e., the summed global and local kernels). Furthermore, the length scale and amplitude posterior distributions for both local kernels are not well-constrained and consistent with zero (Figure \ref{fig:hyperparams}).  These inferences demonstrate that hyperparameter posterior distributions can also diagnose whether a given kernel is needed, while not relying solely on Bayesian model comparisons. In general, the use of more complex kernels with combinations of local and global kernels may therefore be a preferred, less assumptive approach for marginalizing over any possible sub-structure. Unneeded kernels will both be penalized in Bayesian evidence and have hyperparameter posterior distributions consistent with zero.

The local kernels identify two sources of localized data-model inference mismatches; one at 1.18 $\mu$m and one at wavelengths greater than 2.7 $\mu$m. The latter overlaps significantly with the strongest CO$_2$ band in the NIRISS wavelength range, implying that previous inferences of a high CO$_2$ abundance may be affected by an unknown in the data. As such, the wider precision of CO$_2$ abundances in the GP-aided cases better reflects our true inference capabilities from this dataset.

Without repeat observations, it remains difficult to determine the origin of the local kernel placement at 2.7~$\mu$m; whether it reflects a noise fluctuation, a physical mechanism producing a secondary feature, or a higher-than-expected CO$_2$ abundance. Additional NIRISS observations covering the same wavelength range could help clarify this feature's origin. In parallel, we recommend comparing these results with upcoming NIRSpec/G395H observations of WASP-96~b (GO 4082, PI: Radica), which will include the stronger 4.3~$\mu$m CO$_2$ absorption band. Joint constraints from multiple CO$_2$ features from the combined NIRISS and NIRSpec data will improve the robustness and precision of CO$_2$ abundance estimates relative to those derived from NIRISS alone.

The feature at 1.18~$\mu$m corresponds to two adjacent data points lying below the model, both identified as outliers in \citet{radica_awesome_2023}. This same region also presented a challenge in the retrievals of \citet{taylor_awesome_2023}, further suggesting that the feature warrants continued investigation. A re-analysis of HST data from \citet{nikolov_solar--supersolar_2022}, along with future JWST observations covering the $\sim$1--2~$\mu$m range, may help determine the astrophysical or instrumental nature of the feature.

\subsection{Assessing the Impact of Background Star Contamination} \label{sec:contaminants}

\citet{radica_awesome_2023} identified two contaminated regions of the NIRISS spectrum during the data reduction process (see Figure C3 of \citealt{radica_awesome_2023}), which each arose from an overlap of a background stellar artifact with the orders on the NIRISS detector. Due to the slitless nature of the NIRISS/SOSS detector, there is both an ``Order 0" contaminant (an undispersed background star) and an ``Order 1" contaminant (the dispersed spectrum of another background star, overlapping with the spectrum of the planet). The background stars imprinted features on the transmission spectrum at roughly 1.3--1.7 $\mu$m and $\sim$2.1 $\mu$m, with feature sizes of $\lesssim 100$ and 750 ppm, respectively. \citet{radica_awesome_2023} developed a methodology to correct for these contaminants and applied it at the data reduction level.

We include the spectrum with uncorrected background contamination in our analysis, although we find that it does not strongly impact our inferred results. The global kernel is able to marginalize over both the contaminants in the data so that local kernels are not needed at those wavelength ranges. Additionally, the retrieved inferences for atmospheric parameters are consistent between both the corrected and uncorrected data. We therefore conclude that background contamination in the spectrum did not significantly contribute to any biasing effects in the analysis of WASP-96 b. The retrieved spectra and posterior distributions for the dataset with uncorrected contamination can be found in the repository for this manuscript\footnote{\url{https://osf.io/pksnq/}}.

\section{Conclusions and Discussion}\label{sec:discussion}
With the advent of JWST, the precision of exoplanet spectroscopy has dramatically increased, making it critical to ensure that modeling assumptions do not introduce systematic biases into atmospheric retrievals. In this work, we incorporated Gaussian Processes (GPs) directly into a transmission spectrum retrieval framework, enabling it to marginalize over structured residuals that arise from model–data mismatches, including missing absorbers in the model or data unknowns. This approach improves the robustness of inferred atmospheric properties, particularly when dealing with incomplete models or unaccounted-for noise. We summarize our main findings below:

\begin{enumerate}
  \item In the presence of unknown spectral features, missing model physics, or correlated noise, traditional retrieval methods can yield highly precise but inaccurate constraints on atmospheric properties.

  \item Incorporating Gaussian processes into retrieval frameworks may enable more accurate atmospheric inferences, particularly in the presence of unknowns that can be well-approximated by the chosen GP kernels.

  \item GP-aided retrievals can not only improve model fits to the data, but also identify regions of the spectrum where model–data mismatches occur.
  
  \item The bias–variance tradeoff provides a useful diagnostic for evaluating retrieval performance by capturing both accuracy and precision in a single metric. Incorporating variance, alongside bias, offers a more complete view of inference quality.
  
  \item The combined global and local GP framework minimizes the bias–variance tradeoff for synthetic data and achieves the lowest mean squared error across most species, providing an interpretable and balanced posterior distribution in the presence of structured residuals.
  
  \item For H$_2$O, one of the best-constrained molecules in the NIRISS bandpass, our GP-aided retrievals are preferred by Bayesian evidence and yield a broader posterior ($\sim$0.1 dex wider) than traditional methods, suggesting that previous estimates of H$_2$O abundance may be over-precise.

  \item In the case of CO$_2$, the broader constraints ($\sim$1 dex) and higher Bayesian evidence obtained with GP-aided retrievals suggest that current discrepancies in the literature may stem from over-precision rather than true disagreement. The difficulty of constraining CO$_2$ from a single, partially covered absorption band in the NIRISS bandpass likely contributes to this effect.
  
  \item Our GP framework reveals an underlying correlated noise structure across the NIRISS bandpass in observations of WASP-96~b and localizes two regions of significant model–data mismatch, including one overlapping with the CO$_2$ feature at 2.7~$\mu$m. This mismatch may bias the retrieved CO$_2$ abundance when using traditional retrievals that do not account for correlated noise.

\end{enumerate}

We discuss key considerations for implementing Gaussian processes within retrieval frameworks and examine how kernel selection and prior choices can influence the inferred atmospheric properties.

\subsection{Choice of Hyperparameter Models and Priors} \label{sec:hyperparams_discussion}

Selecting an appropriate GP kernel can be challenging, particularly when the nature of the correlated noise is not well understood \textit{a priori}. Ideally, the kernel should approximate the physical processes that give rise to the residual structure in the noise. For instance, time-series observations of stellar variability have been modeled using damped harmonic oscillator kernels \citep{foreman-mackey_fast_2017, pereira_gaussian_2019}, while similar oscillatory kernels have been employed to mitigate stellar effects in transit light curves \citep[e.g.,][]{radica_muted_2024, coulombe_highly_2025}. In the context of exoplanet light curve fitting, both global SE and Matérn-3/2 kernels have been used to capture systematic correlations between adjacent observations \citep[e.g.,][]{gibson_gaussian_2012, evans_uniform_2015, fortune_how_2024}. Likewise, \citet{czekala_constructing_2015} applied both global and local Matérn-3/2 kernels in wavelength space to fit stellar spectral models to observed data.

We adopt a squared exponential (SE) kernel for both the global and local kernels. This choice is motivated by its simplicity and its ability to approximate smooth, correlated structures in the data. For the global kernel, the width of the NIRISS point spread function (PSF) implies that adjacent wavelength bins are correlated in a way that decays with inter-pixel distance \citep{albert_near_2023}, consistent with the type of behavior modeled using Matérn-3/2 kernels in stellar spectra \citep{czekala_constructing_2015}. In the local case, the chosen structure of the kernel makes it ideal for marginalizing over effects in the spectrum that follow a pseudo-Gaussian shape, such as features from absorbers not accounted for within the model or contaminating stellar features. However, this choice of kernel will likely be unable to account for higher order effects, such as three-dimensional effects for tidally locked planets \citep[e.g.,][]{blecic_implications_2017, caldas_effects_2019}.

For local kernels, the squared exponential (SE) kernel is well-suited to approximate unmodeled spectral features in low-resolution data, even without prior knowledge of their exact shape. We compare the performance of the squared exponential and Matérn-3/2 kernels on our synthetic dataset, but find no significant differences in the retrieved atmospheric properties, and therefore adopt the former throughout this work. We do not consider periodic kernels in this work as, to the best of our knowledge, no clear wavelength-periodic structure has been identified in NIRISS spectra to date. We note that both the Matérn-3/2 and SE kernel are realizations of the generalized Matérn kernel equation. This generalized equation can provide a more flexible kernel realization if the additional hyperparameter $\nu$, representing the roughness of the covariance, is solved for. However, we consider this outside the scope of this work.

We further examine the impact of hyperparameter prior (hyperprior) selection on retrieval outcomes. We compare log-uniform and uniform priors for the GP hyperparameters using synthetic datasets. As expected, we find that hyperprior choice does not significantly affect the retrieved atmospheric parameters in most cases. This result is consistent with \citet{chen_how_2018}, who showed that while hyperpriors can influence the best-fit hyperparameters, they have limited impact on the inferred model parameters.


However, we find that hyperprior selection does influence how well the retrieval can constrain information about the correlation in the data. In particular, uniform priors on kernel amplitude and length scale may yield narrower posteriors on the location and strength of local correlations, particularly in low-SNR regimes. In contrast, log-uniform priors allocate greater density to small values and may lead to broader posteriors centered near negligible kernel contributions. This behavior reflects the interplay between the prior volume and the informativeness of the likelihood: when the data only weakly constrain the kernel, the choice of prior becomes non-negligible. We adopt uniform priors throughout to more readily detect and quantify non-zero kernel structures, acknowledging that this choice may bias against the null (no-correlation) hypothesis. To ensure that our prior ranges do not restrict the inference, we select bounds that encompass any physically plausible contribution to the transmission spectrum. For example, we allow kernel amplitudes to vary from zero (no GP contribution) to 10,000~ppm (equivalent to a 1\% transit depth), which is conservatively larger than any expected unknown absorber or contaminant.

We also find that the kernel amplitude and length scale become partially degenerate in two limiting regimes. When the amplitude approaches zero, the kernel contribution vanishes and the likelihood becomes insensitive to the length scale, rendering it unconstrained. Conversely, when the length scale becomes smaller than the typical bin spacing of the data ($\sim0.01~\mu$m), the kernel contribution becomes sharply localized, effectively acting as a spike at a single data point. In this limit, the kernel can mimic additional white noise, which could introduce redundancy with the noise model and limit its identifiability.

The choice of hyperparameter priors can also strongly impact the relative model preference (Bayes factor) of a GP noise model when compared to a traditional white noise model against the data. Bayes factors are the ratio of the Bayesian evidences of two models, each of which can be thought of as the weighted average of the Bayesian likelihood across the prior volume. Their ratio corresponds to the statistical significance of the preference of one model over another \citep[e.g.,][]{trotta_bayes_2008}. By expanding the prior range of the hyperparameters, one can thus inadvertently dilute the Bayesian evidence of the GP noise model even in the case of correlated noise in the data. This effect can lead to selectively preferring a traditional noise model, which will in turn require a more complex physical model to overcome the correlated noise structure without a GP. A balance must therefore be struck between a broad hyperprior that contains the possible hyperparameter values to avoid biasing the kernel inference, and one that is not overly broad so as to dilute the evidence \citep[for further discussion of this effect, see, e.g.,][]{mackay_information_2003, trotta_bayes_2008, llorente_safe_2023}. We consider further exploration of prior selection beyond the scope of this work, but recommend that future studies carefully assess the influence of hyperprior choices and adopt physically motivated ranges consistent with the observational setup and scientific goals.

\subsection{Future Directions}
\label{sec:future}

In addition to the detection of new opacity sources, the effects of stellar heterogeneity have emerged as a key challenge in interpreting JWST spectra across the exoplanet population \citep[e.g.,][]{fu_water_2022, lim_atmospheric_2023, fournier-tondreau_near-infrared_2023, may_double_2023, moran_high_2023, fournier-tondreau_transmission_2024, radica_promise_2025}. Accounting for these effects is essential, as uncorrected stellar contamination can bias inferences of atmospheric composition and structure \citep{iyer_influence_2020}. However, stellar contamination remains difficult to model, in part because stellar atmosphere models often fail to reproduce observed spectra \citep{czekala_constructing_2015, iyer_sphinx_2023}. As a result, traditional corrections may introduce errors when applied to contaminated transmission spectra.

GP-aided retrievals offer a promising alternative by flexibly capturing residual features that standard stellar models cannot. In particular, a GP applied directly to the stellar component within the contamination term (e.g., \citealt{pinhas_retrieval_2018}) could help isolate and marginalize over mismatches between modeled and observed stellar signals. While this application is beyond the scope of the present work, we encourage future studies to explore comparisons between traditional methods to mitigate the effects of stellar heterogeneities \citep[i.e., the Transit Light Source Effect, e.g.,][]{rackham_transit_2018} and GP-based approaches.

Previous work has also explored methods to analytically estimate the spectral covariance matrix directly from the data \citep[e.g.,][]{holmberg_exoplanet_2023, fortune_how_2024}. While such approaches can characterize noise structure without imposing model assumptions, they are mathematically complex, and do not directly address biases introduced by model deficiencies. However, comparing the analytically derived and retrieval-inferred covariance matrices may help disentangle model–data degeneracies. Additionally, the relative simplicity of the inferred covariance matrix from the retrieval allows for scalability to multiple spectra, providing an opportunity for identification of trends from instrumental or physical effects. Although a comparison between the methods is beyond the scope of the present work, future efforts integrating both methodologies could yield valuable insights into the limitations of current retrieval frameworks.

\subsection{Concluding Remarks} \label{sec:conclusions}

As we move toward a population-level understanding of exoplanet formation and evolution in the coming decades, it is essential that our inferences from individual systems are both accurate and trustworthy. This work highlights several key findings, including the risk of overly precise, but ultimately biased, inferences when retrievals are applied to contaminated data or rely on incomplete models. We show that such limitations can be mitigated with GP-aided retrievals, which offer a flexible toolkit for identifying and marginalizing over hidden sources of uncertainty.

These uncertainties may originate from the data itself or from assumptions embedded in the model. Understanding their origin is critical to avoiding biases that could propagate through comparative studies and misinform our interpretation of exoplanet populations. By recognizing the limitations of both current models and the data, including the effects of correlated and systematic noise, we are better positioned to extract robust constraints from JWST observations and fully capitalize on the capabilities of future missions in the decades ahead.
\\
\\
We thank the referee for thoughtful comments that significantly improved this work. Y.R. thanks Dr. Aishwarya Iyer for meaningful conversations during the early stages of this work. This research was enabled thanks to support from NASA XRP grant 80NSSC24K0160. L.W. thanks the Heising-Simons Foundation for their funding through the 51 Pegasi b Postdoctoral Fellowship, and Arizona State University for their support through the Presidential Postdoctoral Fellowship Program.  Y.R.,\ L.W.,\ and M.R.L.\ acknowledge Research Computing at Arizona State University for providing HPC and storage resources that have significantly contributed to the research results reported within this manuscript. This work was performed under the auspices of the U.S. Department of Energy by Lawrence Livermore National Laboratory under Contract DE-AC52-07NA27344. The document release number is LLNL-JRNL-2003715.

\newpage
\appendix
\renewcommand{\thefigure}{A\arabic{figure}}
\setcounter{figure}{0}

\section{Synthetic Model with a Higher-Amplitude Feature}\label{app:feature}

Here, we show the same spectrum as described in Section \ref{sec:tests}, but with an injected feature with a higher amplitude, while keeping all other parameters and hyperparameters consistent. We retrieve on this synthetic dataset using our traditional, global, and combined global and local GP frameworks. Similarly to the other spectrum, our GP-aided frameworks are able to provide more accurate estimates of both key absorber abundances and the kernel hyperparameters. 

\begin{figure*}[h]
    \centering
    \includegraphics[width=\textwidth]{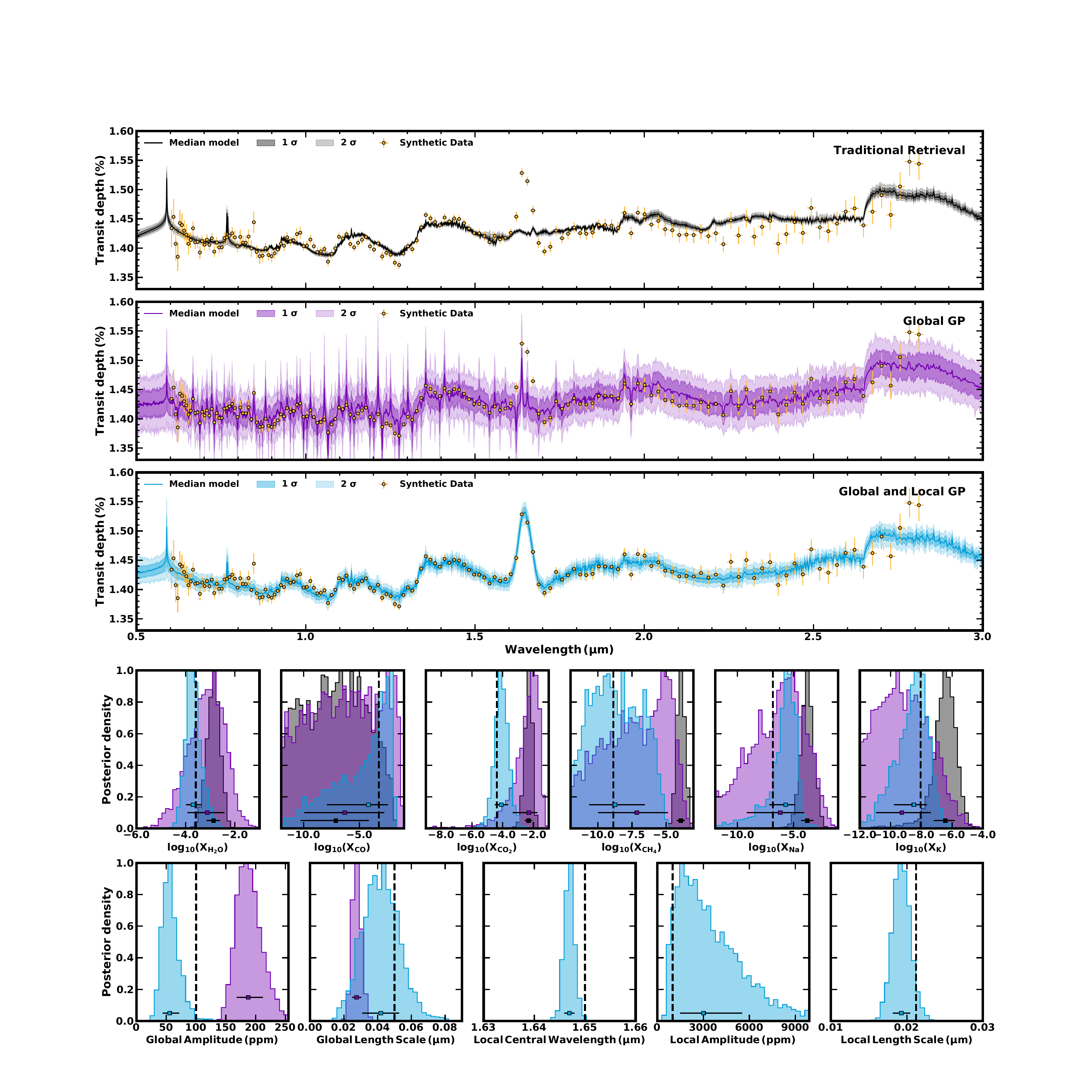}
    \caption{The traditional, global, and combined global and local kernel retrievals for the synthetic dataset with a larger feature, as well as the retrieved absorber abundances for each case. The combined global and local kernel retrieval provides the most accurate inferences, while marginalizing over the injected feature at 1.65 $\mu$m. The retrieved hyperparameters are shown below. Again, we find that the combined global and local kernel retrieval provides the largely accurate inferences, although it slightly underestimates the amplitude of the global kernel.}
    \label{fig:bigfeature}
\end{figure*}

In this case, the global GP is again preferred over a traditional retrieval, this time by 17.3$\sigma$ ($\ln B=146.55$); however, the combined local and global retrieval is preferred over the global retrieval by 7.1$\sigma$ ($\ln B=23.29$), implying a strong need for a local kernel. Thus, we conclude that the lack of strong ``detection" of the local kernel in Section \ref{sec:tests} is driven by the small amplitude of the feature, rather than the inefficiency of a local kernel. The retrieved spectra, absorber abundances, and hyperparameters are shown in Figure \ref{fig:bigfeature}.

\section{WASP-96 b Retrieved Parameters}\label{app:summary_table}

This appendix provides the full set of retrieved parameter values for each retrieval configuration described in the main text: traditional (T), global GP only (G), and global GP with one (G1L) and two (G2L) local kernels. These values are presented in Table~\ref{table:retrieved} and include molecular abundances, temperature profile parameters, cloud and haze properties, and GP hyperparameters. The table complements the summary figures in Section~\ref{sec:W96}.

\begin{deluxetable}{lcccc}
\tablecaption{Retrieved parameter estimates ($\pm1\sigma$) for each retrieval configuration: traditional (T), global GP only (G), and global GP with one or two local kernels (G1L, G2L). Units match those in Table~\ref{table:W96_prior}.\label{table:retrieved}}
\tablehead{
\colhead{\textbf{Parameter}} & \colhead{\textbf{T}} & \colhead{\textbf{G}} & \colhead{\textbf{G1L}} & \colhead{\textbf{G2L}}
}
\startdata
\cutinhead{Molecular Abundances}
$\log_{10}\text{X}_{\text{H}_2\text{O}}$ & $-3.66^{+0.39}_{-0.32}$ & $-3.60^{+0.49}_{-0.42}$ & $-3.45^{+0.51}_{-0.44}$ & $-3.51^{+0.53}_{-0.43}$ \\
$\log_{10}\text{X}_{\text{CO}}$         & $-5.20^{+2.15}_{-4.01}$ & $-6.12^{+2.61}_{-3.57}$ & $-5.21^{+2.24}_{-3.86}$ & $-5.77^{+2.68}_{-3.69}$ \\
$\log_{10}\text{X}_{\text{CO}_2}$       & $-4.63^{+0.48}_{-0.50}$ & $-4.63^{+0.59}_{-0.65}$ & $-4.43^{+0.55}_{-0.57}$ & $-4.87^{+0.80}_{-2.21}$ \\
$\log_{10}\text{X}_{\text{CH}_4}$       & $-8.64^{+2.11}_{-2.15}$ & $-8.43^{+2.02}_{-2.07}$ & $-8.25^{+2.13}_{-2.25}$ & $-8.49^{+2.12}_{-2.08}$ \\
$\log_{10}\text{X}_{\text{NH}_3}$       & $-8.88^{+1.94}_{-1.98}$ & $-8.63^{+1.94}_{-2.09}$ & $-8.79^{+2.09}_{-2.01}$ & $-8.69^{+2.11}_{-2.04}$ \\
$\log_{10}\text{X}_{\text{HCN}}$        & $-8.11^{+2.67}_{-2.52}$ & $-7.77^{+2.40}_{-2.48}$ & $-7.88^{+2.40}_{-2.51}$ & $-7.89^{+2.53}_{-2.54}$ \\
$\log_{10}\text{X}_{\text{Na}}$         & $-8.42^{+1.34}_{-2.22}$ & $-8.72^{+1.44}_{-1.81}$ & $-8.61^{+1.54}_{-1.95}$ & $-8.79^{+1.42}_{-1.89}$ \\
$\log_{10}\text{X}_{\text{K}}$          & $-7.41^{+0.97}_{-1.76}$ & $-8.20^{+1.56}_{-1.44}$ & $-8.35^{+1.92}_{-1.26}$ & $-8.53^{+1.60}_{-1.18}$ \\
\cutinhead{Temperature Profile}
$T_0$ [K]                             & $1058^{+132}_{-120}$ & $1036^{+157}_{-125}$ & $1024^{+144}_{-120}$ & $1027^{+149}_{-130}$ \\
$\log_{10}P_{1}$ [bar]               & $-2.90^{+1.86}_{-2.21}$ & $-2.88^{+1.71}_{-1.90}$ & $-2.80^{+1.75}_{-2.06}$ & $-2.84^{+1.74}_{-2.03}$ \\
$\log_{10}P_{2}$ [bar]               & $-5.90^{+2.25}_{-1.82}$ & $-5.93^{+2.15}_{-1.70}$ & $-5.95^{+2.11}_{-1.73}$ & $-6.03^{+2.06}_{-1.68}$ \\
$\log_{10}P_{3}$ [bar]               & $-0.16^{+0.92}_{-1.10}$ & $-0.17^{+0.87}_{-0.95}$ & $-0.15^{+0.87}_{-1.01}$ & $-0.22^{+0.87}_{-1.00}$ \\
$\alpha_1$ [K$^{-1/2}$]              & $1.45^{+0.34}_{-0.39}$ & $1.48^{+0.33}_{-0.34}$ & $1.43^{+0.35}_{-0.37}$ & $1.45^{+0.34}_{-0.35}$ \\
$\alpha_2$ [K$^{-1/2}$]              & $1.39^{+0.40}_{-0.46}$ & $1.35^{+0.41}_{-0.42}$ & $1.33^{+0.41}_{-0.50}$ & $1.36^{+0.39}_{-0.44}$ \\
\cutinhead{Clouds and Hazes}
$\log_{10} \kappa_{\rm cloud}$ [m$^2$] & $-35.68^{+2.83}_{-2.82}$ & $-35.49^{+2.86}_{-2.77}$ & $-35.34^{+2.86}_{-2.94}$ & $-35.17^{+2.88}_{-3.01}$ \\
$\phi_{\rm cloud}$                   & $0.92^{+0.06}_{-0.13}$ & $0.69^{+0.19}_{-0.18}$ & $0.67^{+0.20}_{-0.19}$ & $0.62^{+0.19}_{-0.16}$ \\
$\log_{10}a$                         & $1.81^{+0.53}_{-0.33}$ & $2.68^{+1.17}_{-0.91}$ & $2.92^{+1.19}_{-0.95}$ & $3.04^{+1.13}_{-0.90}$ \\
$\gamma$                             & $3.56^{+0.83}_{-0.60}$ & $4.84^{+2.07}_{-1.51}$ & $4.87^{+1.96}_{-1.41}$ & $5.13^{+1.99}_{-1.42}$ \\
\cutinhead{GP Hyperparameters}
$a_G$ [ppm]                         & -- & $86.63^{+12.05}_{-11.05}$ & $73.36^{+11.93}_{-11.11}$ & $71.84^{+12.35}_{-11.06}$ \\
$L_G$ [$\mu$m]                      & -- & $0.012^{+0.003}_{-0.003}$ & $0.011^{+0.004}_{-0.004}$ & $0.012^{+0.004}_{-0.004}$ \\
$\mu_1$ [$\mu$m]                    & -- & -- & $1.18^{+0.003}_{-0.003}$ & $1.18^{+0.003}_{-0.004}$ \\
$a_{1}$ [ppm]                     & -- & -- & $1583^{+2918}_{-1124}$ & $1533^{+2623}_{-1093}$ \\
$L_{1}$ [$\mu$m]                  & -- & -- & $0.014^{+0.007}_{-0.005}$ & $0.014^{+0.007}_{-0.005}$ \\
$\mu_2$ [$\mu$m]                    & -- & -- & -- & $2.718^{+0.131}_{-0.647}$ \\
$a_{2}$ [ppm]                     & -- & -- & -- & $2494^{+3215}_{-1912}$ \\
$L_{2}$ [$\mu$m]                  & -- & -- & -- & $0.021^{+0.061}_{-0.011}$ \\
\enddata
\end{deluxetable}

\bibliography{bibliography_all}{}
\bibliographystyle{aasjournal}

\end{document}